\PassOptionsToPackage{hyphens}{url}
\documentclass[sigconf,nonacm,10pt]{acmart}

\setcitestyle{nocompress}
\makeatletter
\AtBeginDocument{\let\NAT@space\@empty}
\makeatother

\usepackage{microtype}
\usepackage{booktabs}
\usepackage{tabularx}
\usepackage{array}
\usepackage{ragged2e}
\graphicspath{{figures/}}

\makeatletter
\patchcmd{\@mkauthors@iii}
  {\and\par\bigskip}
  {\and\par\bigskip\vspace{6pt}}
  {}{\PackageError{orphan}{Author spacing patch failed}{}}
\makeatother

\AtBeginDocument{%
  \fancypagestyle{standardpagestyle}{\fancyhf{}%
    \fancyfoot[C]{\fontsize{10}{12}\selectfont\thepage}}%
  \fancypagestyle{firstpagestyle}{\fancyhf{}%
    \fancyfoot[C]{\fontsize{10}{12}\selectfont\thepage}}%
  \pagestyle{standardpagestyle}%
  \raggedbottom
  \setlength{\footskip}{30pt}}

\newcolumntype{Y}{>{\RaggedRight\arraybackslash}X}
\AtBeginEnvironment{thebibliography}{\RaggedRight\setlength{\bibsep}{0pt plus 0.3pt}}
\apptocmd{\thebibliography}{\clubpenalty=0 \widowpenalty=0 \interlinepenalty=0}{}{\PackageError{orphan}{Bibliography penalty patch failed}{}}

\AtBeginDocument{\captionsetup[figure]{skip=4pt}\captionsetup[table]{skip=6pt}}

\title{Extending DBLog-Based Incremental Snapshots to Handle Orphan Rows in
CDC Pipelines}
\author{Andreas Andreakis}

\begin{document}

\begin{abstract}
Applications use heterogeneous datastores where one is the source and the others are sinks that serve specific needs as derived copies, such as a search index or data warehouse. To keep sinks up-to-date, change data capture (CDC) is standard practice to propagate ongoing source changes. However, as CDC relies on a transaction log with limited retention, it cannot by itself bootstrap sinks with the full data set, and traditional approaches either stall source writes or hold back log events. DBLog-based incremental snapshots, as adopted by Debezium and Flink CDC, solve this by reading source tables in chunks interleaved with the log, without blocking writes or stalling log events for prolonged periods, and allow reads to be interrupted and resumed where they stopped.

After the initial bootstrap, a sink can drift when losing change events or receiving unintended writes. Even repeating a bootstrap can still leave orphan rows that exist in the sink and no longer at the source. A common remedy is to create and populate a new empty sink. Repairing an existing sink can be a better option because it avoids running parallel pipelines, coordinating a switch to a new sink and serving traffic from a drifted sink until the new sink is fully ready.

In DBLog's original design, each chunk only emits the latest state of the rows the table still contains, so it does not handle orphan rows. In this work we extend DBLog to also delete orphan rows at sinks. We use what a source chunk read already knows, namely the keys that the key range select did not return, emitting a new kind of event that sinks apply as deletes to remove orphans. We also add a repair request that an external process can call with orphan candidates, which the connector checks with its own read, emitting these events only for keys confirmed absent at the source. This enables three modes. In the fully integrated mode, every chunk pass removes the orphans it finds, which suits both full and periodic proactive repairs. In the second mode, an external drift process hands suspected orphans at the sink to the connector, which deletes only those confirmed absent. The third mode combines both and is what we recommend.

Together, these extensions turn DBLog into a complete data synchronization engine whether a sink catches up on live changes or missing rows (empty sink or lost inserts), holds wrong column values (corrupted data or lost updates), or contains rows it should not have (orphan rows). Their core safety and convergence results have been formally verified in Isabelle/HOL and Lean~4.
\end{abstract}

\maketitle

\section{Introduction}\label{sec:intro}

It is common to see applications that store the same data across multiple datastores, each serving
a specific need~\cite{kleppmann2019olep,laigner2021microservices}. For example, a relational database
may hold the canonical data while a search index handles full-text queries and a
data warehouse provides analytics. In such environments, each downstream sink needs to
stay synchronized as a secondary copy of the source database. Change data capture
(CDC) has become a widely used method to keep these sinks current~\cite{kleppmann2019olep,andreakis2020dblog}. A CDC connector
reads the database's transaction log and publishes every committed insert,
update and delete as a change event, which each sink applies in order shortly
after the source changes.

However, the transaction log alone cannot populate a new sink. Because databases
retain their transaction logs for a limited time or size, the log usually does not hold the
complete history of changes. A sink therefore needs an initial full copy of the
source data, a bootstrap, before it can follow ongoing changes. Full or partial
copies are also needed later on, for instance when a source database is restored
from a backup or when downstream data is lost or corrupted and needs
repair~\cite{andreakis2020dblog}.
Taking that copy while changes continue to arrive is tricky because copied rows
and live change events must be ordered so that an older value never overwrites a
newer one. Traditional snapshot approaches copy a table in a single long read,
often acquiring table locks that block application writes (at least briefly) or 
pausing log processing until the read finishes and is applied to the sinks. 
They must also usually start over from the beginning if interrupted.

DBLog~\cite{andreakis2020dblog} copies a table while the change stream keeps
flowing. It selects rows from the table in chunks ordered by primary key
and brackets each chunk read between two markers that are written to the transaction log,
a low and a high watermark. Any row in the chunk whose key also changed in the
log between the two watermarks is dropped from the chunk in memory. This happens
because the chunk query executes somewhere between the two watermarks and since
its exact log position is unknown, DBLog relies on the log events for any
overlapping keys instead. When the high watermark
arrives in the log, DBLog emits the remaining chunk rows into its output stream ahead
of any subsequent changes. No tables are locked and a snapshot can be triggered
at any time for an entire table or for specific keys. Because progress is
tracked per chunk, a snapshot can also pause and resume where it stopped.
Debezium~\cite{debezium_16_final,debezium_incremental_snapshots} and Flink
CDC~\cite{flinkcdc_mysql} adopted this watermark-based design for their
incremental snapshots and our recent work shows that all of these
variants belong to the same family~\cite{andreakis2026generalized}. Section~\ref{sec:chunk}
walks through this loop step by step and the original DBLog paper and Netflix
Technology Blog post describe it in further detail~\cite{andreakis2020dblog,dblog_netflix_blog}.

No matter which approach is used, a bootstrapped sink can later drift from the source in
ways the change stream cannot repair. Consider an \texttt{orders} table with
primary keys 1 to 10 that is replicated by a CDC pipeline to a search index.
Suppose the connector is down for longer than the database retains its
transaction log and during that outage the application deletes orders 4 and 7.
When the connector restarts, it detects that the source log was truncated.
Because the team cannot tell which change events the search index missed, they
trigger a fresh incremental snapshot. The snapshot copies the eight remaining
orders and streaming resumes, yet the search index is still out of sync
because orders 4 and 7 remain in the index.

A row that exists in a sink but not at the source is an \emph{orphan row}. The truncated transaction log no longer contains the two
deletes and running a new snapshot only refreshes the rows that still exist at
the source. In DBLog's original design, each chunk emits the latest state of the
rows present in the table, with no built-in ability to detect and delete orphans downstream. 
Debezium's and Flink CDC's incremental snapshots behave the same way.

Lost deletes are only one cause of orphans. A sink can also accumulate extra rows when an external process writes to it 
directly (Section~\ref{sec:kinds}). Because sinks can drift when data loss or corruption occurs in ways that existing monitoring may
not capture, teams that experience such issues may also want to run periodic repairs to catch drift early.

A common workaround is to provision a new, empty sink, bootstrap it from scratch,
and switch application traffic over once the load completes. We believe, however, 
that repairing the existing sink in place through the running pipeline can be a better option.
In-place repair avoids operating two parallel pipelines and coordinating a
cutover to a new datastore. Bootstrapping a large sink also takes time, during
which the application must continue serving traffic from the stale or corrupted
sink until the replacement is fully ready.

\begin{figure*}[t]
\centering
\includegraphics[trim=0 2.5pt 0 2.5pt,clip]{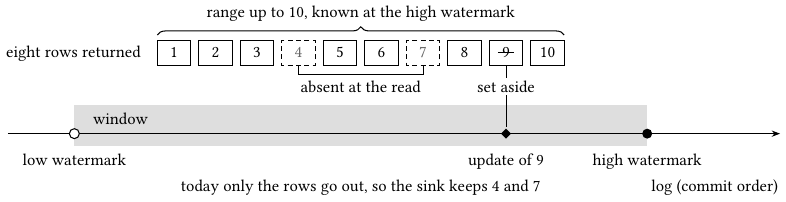}
\caption{One chunk read between two watermarks, over the key range up to 10,
with order 9 updated inside the window. The read returns eight rows and shows
that 4 and 7 do not exist, for a known range at a known position. Today only
the rows go out, so a sink that holds orders 4 and 7 keeps them.}
\label{fig:read}
\end{figure*}

In the original DBLog algorithm, when the snapshot's first chunk read returned orders 1, 2,
3, 5, 6, 8, 9 and 10, that query also proved that orders 4 and 7 do not exist
within the queried key range (Figure~\ref{fig:read}). The original DBLog design emits the present rows and ignores this second half 
of the query result. The core insight of this paper is that the chunk read already captures the missing
information. In other words, a pass of chunk reads that uses the absent keys, combined with
the ongoing log stream, carries everything needed to bring a sink back into sync
with the source across any insert, update, or delete drift while the pipeline keeps running.

In this work, we extend DBLog-based incremental snapshots with three components.
First, we introduce the \emph{chunk info record}, a metadata record that the
connector emits alongside each chunk's rows. This record specifies the chunk's
key range and two key lists. The \emph{selected keys} are the keys the query
returned. The \emph{changed keys} are the keys in the range with a change event
inside the watermark window.
A sink connector that accepts chunk info records applies all preceding
stream events first, deletes every key the sink holds within the chunk's range
that is neither selected nor changed (in the example removing orders 4 and 7),
and then applies the chunk's rows and subsequent events. For key-partitioned transports such as Kafka topics and sinks that only
accept per-key events, the source connector can instead render the chunk info record into
per-key \emph{absence events}, one per absent candidate key supplied to it,
which sinks apply as deletes.
Second, we add a connector repair request named
\texttt{verify-keys-absent}. An external process submits keys it suspects are orphans. The connector queries those 
keys at the source between watermarks and emits a positioned absence event into the stream for each key confirmed absent. The requesting 
process never deletes from the sink directly.
Third, we pair the connector with an external \emph{drift process} that inspects
sinks for suspected orphans and submits them via the new repair request. We define
this process by the safety contract it must satisfy rather than prescribing a
single implementation.

These components support three operating modes. In the fully integrated mode,
every chunk-driven snapshot pass removes the orphans it identifies, which suits
both full-table repairs when the scope of corruption is unknown and routine
periodic repairs. In the second mode, the external drift process monitors the
sinks and submits suspected orphans to the connector via the repair request,
catching orphans and other drift faster than periodic full scans alone. The
third mode runs both together to combine their strengths, which is the setup we
recommend.

Together, these extensions turn a DBLog-based pipeline into a complete data
synchronization engine. While live changes continue to flow, a snapshot pass already fills in missing rows (whether
from an empty start or lost inserts) and overwrites wrong column values (whether
from downstream corruption or lost updates), because each chunk emits the latest
state of every present row. By using the absent keys from each chunk read,
the same pass can now also remove orphan rows that the sink should not hold. 

We have formally verified the safety and convergence of these extensions in
Isabelle/HOL on top of the Generalized DBLog contract
core~\cite{andreakis2026generalized,generalized_dblog_artifact}. For any sink
that applies the stream in order from an arbitrary initial state, statements of
absence never produce a wrong value or delete a live row and the sink converges
to the exact source state if and only if those statements cover every orphan the
sink holds that the log stream never mentions, under the assumptions detailed in Section~\ref{sec:checked}.

Our scope throughout this paper is row-level data synchronization over committed
inserts, updates and deletes, while schema changes and other non-row operations
are outside the scope of this work.

This paper is written for teams operating CDC pipelines and for maintainers of CDC connectors 
such as the Debezium community~\cite{debezium_16_final,debezium_incremental_snapshots,andreakis2026generalized}. 
Sections~\ref{sec:chunk} and~\ref{sec:gap} review DBLog's chunk read and show what each read establishes and what it
currently discards. Section~\ref{sec:kinds} examines how orphan rows arise in
production and how teams detect them today. Section~\ref{sec:mechanisms}
presents our extensions and their three operating modes.
Sections~\ref{sec:diff} and~\ref{sec:compare} analyze the common baseline of an
external diff-and-delete job, state the ordering condition its deletes must
satisfy and compare the tradeoffs of the three approaches. Finally, we cover
the formal verification in Isabelle/HOL (Section~\ref{sec:checked}), prior art
(Section~\ref{sec:prior}) and conclusions (Section~\ref{sec:future}).

\section{The chunk read}\label{sec:chunk}

DBLog interleaves chunked table selects with the transaction log in a continuous
loop~\cite{andreakis2020dblog,andreakis2026generalized}. The connector reads
committed inserts, updates and deletes from the source log and publishes them
to an ordered output event stream that downstream sinks apply in sequence.
Throughout this paper, ``the connector'' means this source connector. We call
the component that applies the stream to a sink the sink connector. A sink is
any consumer of the stream, such as a database, a search index, a cache, or even a
service behind an API. The connector never reads a sink and only emits
events into its output stream. Whether and how an event is applied is up to each sink.
To capture a table's state, the connector
selects rows in chunks of bounded size ordered by primary key, where each chunk
select starts strictly after the last primary key returned by the previous chunk
(including composite primary keys ordered lexicographically across key columns).
For each chunk, the connector briefly pauses log event processing, writes a low
watermark into the transaction log by updating a dedicated watermark table at the source,
executes the chunk \texttt{SELECT}, writes a high watermark and resumes log
processing. The log segment between the low and high watermarks forms the
chunk's \emph{window}. As log events inside the window are processed, any row in
the buffered chunk whose key appears in the window is removed from the chunk. 
This happens because the log event carries the latest state for that key and the 
chunk query may have selected an older state. Finally, when the high
watermark event is reached in the log, which we call the chunk's \emph{close},
the surviving chunk rows are emitted into the stream at that exact log position.

Debezium provides on-demand
incremental snapshots that an operator or tool can trigger at any time while
streaming continues by writing a command to a control table in the source
database (a signaling table) or to a control Kafka
topic~\cite{debezium_incremental_snapshots,debezium_signalling}. These
incremental snapshots follow the same watermark
loop~\cite{debezium_incremental_snapshots,debezium_mysql_connector}, writing
open and close markers to the signaling table in place of the watermark table.
Debezium also supports a read-only mode, using transaction identifiers read from the
database as the watermarks. In that case the connector closes the window once the log advances past the
high-watermark transaction set~\cite{debezium_readonly_blog,debezium_postgresql_connector}
and emits the surviving chunk rows as read events, which sinks apply as upserts. Flink CDC's incremental
snapshot framework records low and high log offsets around each chunk read,
upserts the window's log changes directly into the buffered chunk and emits the
merged rows as insert events~\cite{flinkcdc_mysql}. In all of these
implementations, if a key within the chunk's range is neither returned by the
query nor changed inside the window, the connector emits nothing for it.

\section{What one chunk read establishes}\label{sec:gap}

Each chunk query covers a well-defined key range. In the typical case, that
range runs from the previous chunk's last key to the current chunk's last key. 
A bracketed chunk read establishes three facts at the source. First, it identifies the rows present in
that key range at the moment of the query, which the connector emits after
removing any keys changed inside the watermark window. Second, it establishes 
that no other keys existed in that range as the query saw the source. Third, because
the query is bracketed by two watermarks, its result for every key not changed in the window 
takes a known log position at the window's close.

In our running example, suppose the chunk size is eight rows and the first chunk
query returns orders 1, 2, 3, 5, 6, 8, 9 and 10, covering the key range up to
10, while order 9 is updated inside the window. The query provides information
about all ten keys up to 10. However, the existing design emits the seven
unchanged rows (1, 2, 3, 5, 6, 8 and 10), leaves key 9 to the log stream and discards the fact that keys 4 and 7 were absent (Figure~\ref{fig:read}).

Ignoring absent keys works when bootstrapping an empty sink. If every row in
the sink arrived through the stream in order, any key the stream never mentions
is naturally absent at the sink as well. That is the setting covered by the
Generalized DBLog paper~\cite{andreakis2026generalized}, which notes that
repairing an already populated sink cannot rely on omission and must instead
either clear the repaired key range beforehand or materialize the source's
absences as delete records (tombstones) for the missing keys.

This boundary is visible in existing connectors. Debezium's connector
documentation states that an incremental snapshot emits a read event for each
row it captures~\cite{debezium_mysql_connector,debezium_postgresql_connector}. 
The PostgreSQL connector reference notes that the snapshot flag is always
false for delete events~\cite{debezium_postgresql_connector}. Because a source
connector only reads the source database and has no visibility into downstream
sinks, reconciling extra rows at a drifted target has traditionally been left to
snapshotting into a new table or running a separate sink-side tool.
Section~\ref{sec:mechanisms} shows how a source connector can provide the
missing delete information from its own reads without ever querying the sink.

\section{Where orphans come from and how they are found}\label{sec:kinds}

Table~\ref{tab:kinds} summarizes five ways a sink can end up holding rows the source does not have, 
along with what information remains in the pipeline
(the connector and its output topics) afterward. For three of these causes,
namely a restored row, a skipped record and a forgotten tombstone, the
connector did emit a delete event when the row was removed at the source, so
what the pipeline still knows depends on whether that delete record is still
retained. In Kafka, a delete is represented in a compacted topic by a
\emph{tombstone}, a message with a key and a null payload. Log compaction
retains at least the latest record for each key, but tombstones become eligible 
for removal once the configured tombstone retention period expires, which is 24 hours by default~\cite{kafka_docs}.

\begin{table*}[t]
\caption{Five causes of orphans, how each arises and what the pipeline still knows afterwards.}
\label{tab:kinds}
\footnotesize
\begin{tabularx}{\textwidth}{@{}>{\RaggedRight}p{1.2in}YY@{}}
\toprule
Cause & How it arises & What the pipeline still knows \\
\midrule
Lost delete & The connector is down for longer than the source retains its
transaction log, which MySQL purges after an expiry
period~\cite{mysql_binlog_expire} and a PostgreSQL replication slot retains
until dropped or invalidated by a configured limit~\cite{postgres_replication_slots}. & The keys and their last known values
in the connector's compacted topics. \\
Restored row & The sink is restored from a backup taken before certain deletes
were applied. & The tombstones in the topic, while tombstone retention lasts. \\
Skipped record & A sink consumer fails on a delete record and skips it, or
routes it to a dead-letter queue that is never
replayed~\cite{kafka_connect_configs}. & The delete record in the topic, while
tombstone retention lasts. \\
Side writer & A backfill job, migration script, or manual fix writes rows
directly into the sink that never existed at the source. & Nothing, because the
connector never emitted these keys. \\
Forgotten tombstone & A consumer lags for longer than the topic's tombstone
retention period and misses a tombstone removed by compaction. & Nothing in the
topic, though the source still shows the key absent. \\
\bottomrule
\end{tabularx}
\end{table*}

Teams today rely on three common techniques to detect drift. Comparing row
counts between source and sink serves only as a coarse alarm. Reading and
comparing primary keys from both datastores is exact during a quiet window but
requires full scans of both sides. Computing checksums over key
ranges~\cite{percona_pt_table_checksum} reduces network transfer and catches column value drift as well. 
A fourth source of
information is the incremental snapshot's chunk read itself. Each chunk read
already determines which keys exist in its range at a known log position and using
the absent half of that result turns a snapshot into a single-pass repair that
requires no separate detection scan at the sink.

\section{Extending the design}\label{sec:mechanisms}

\begin{figure*}[t]
\centering
\includegraphics[trim=0 2.5pt 0 2.5pt,clip]{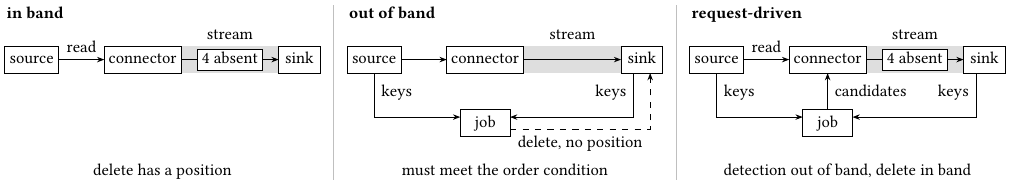}
\caption{Three ways to execute the deletes. In band, the
connector is the coordinator, its read establishes the absence and the delete takes a position
in the stream. Out of band, a job reads both sides and deletes at the sink
with no position, so it must meet the order condition of
Section~\ref{sec:diff}. Request-driven, the job's differences go to the connector in a repair request and the connector's bracketed read gives each confirmed delete its position.}
\label{fig:styles}
\end{figure*}

Deletes that remove orphan rows can be coordinated in two basic ways, or in a
third way that combines them (Figure~\ref{fig:styles}). \emph{In band}, the source connector coordinates the repair. It establishes which
keys are absent through its own bracketed read and places the resulting deletes
into the change stream at that read's close, ensuring that any subsequent
changes follow the delete in stream order and win. \emph{Out of band}, an external process compares the source and sink directly and
issues deletes against the sink outside the stream. Because those deletes have
no position in the stream, they must satisfy the ordering condition in
Section~\ref{sec:diff} to avoid deleting a key that was concurrently re-created
at the source. With a \emph{request-driven} repair, an external process detects
suspected orphans at the sink and submits them to the source connector via a repair
request, named \texttt{verify-keys-absent}, and the connector's own
bracketed read turns each confirmed absence into a positioned absence event in
the stream (Section~\ref{sec:request}). This section details the core mechanism
shared by our extensions (Section~\ref{sec:shared}), the chunk info record (Section~\ref{sec:chunkinfo}), its per-key rendering as absence events
(Section~\ref{sec:absence}), the repair request and sidecar bridge pattern
(Section~\ref{sec:request}), the external drift process and three operating
modes (Section~\ref{sec:drift}), and operational considerations and engine subtleties
(Section~\ref{sec:subtleties}).

\subsection{Shared ground}\label{sec:shared}

We call the connector's output that certifies keys absent a \emph{statement of
absence}. It has two forms, the chunk info record of Section~\ref{sec:chunkinfo}
and the per-key absence events of Section~\ref{sec:absence}, and the repair
request of Section~\ref{sec:request} produces the second form. Every statement
of absence rests on the same invariant. At the moment a bracketed read executes,
every key in its queried scope that the read did not return was absent from the
source as seen by that query. A read narrowed by any row filter beyond its primary-key
range makes no statement of absence, because a row that fails the filter is
omitted from the query result even though it still exists at the source. This
rules out the \texttt{additional-conditions} filter of an on-demand Debezium
snapshot. Both forms then apply the same window exclusion that DBLog
already applies to returned rows. Any
\emph{changed} key (a key with a change event between the low and high
watermarks, including both the old and new keys of a primary-key update) is set aside, meaning its chunk row is dropped and
no statement of absence is made for it, because the log stream carries the latest state for that key. In our running example, 
when order 9 is updated inside the window, its chunk row is dropped just as in standard DBLog and both forms also exclude key 9 
from any statement of absence.

Every statement of absence is emitted at the close of the bracketed read and never at any other stream position. That read is either a
chunk query over a primary-key range or the bracketed key-list query of
Section~\ref{sec:request}, executed between its own watermarks over keys chosen
by the connector or submitted by an external request.
Section~\ref{sec:subtleties} covers the operational considerations and engine
subtleties around Flink CDC's range-merge adaptation, the chunk query's
visibility of logged transactions, key ordering, tail-read limits, safety
caps, table-level aborts and consumer replay.

Before detailing each form, note that a sink connector can use either one or
neither. The chunk info record (Section~\ref{sec:chunkinfo}) fits a consumer of
one ordered stream that can delete the keys it holds in a range, whereas
per-key absence events (Section~\ref{sec:absence}) fit key-partitioned
transports such as Kafka topics and off-the-shelf sink connectors that accept
only per-key deletes.

\subsection{Chunk info record}\label{sec:chunkinfo}

\textbf{Fields.} A chunk info record identifies the table, the chunk's key
range as lower and upper bounds (each marked as inclusive, exclusive, or open),
the selected keys, the changed keys and the identifier of the read. The selected keys are the keys the chunk query
returned. The changed keys are the keys in the range with a change event inside
the window. The log event for such a key carries its latest state, which may be
a delete. A key can be on both lists, as key 9 is in the example that follows. For the first chunk of our \texttt{orders} example,
where keys 1, 2, 3, 5, 6, 8, 9 and 10 were returned and key 9 was updated
inside the window, the chunk info record carries the following logical payload.\par
\begin{footnotesize}
\begin{trivlist}\item\relax
\noindent\begin{minipage}[t]{\columnwidth}
\begin{verbatim}
{
  "table": "inventory.orders",
  "read_id": "snap-01-chunk-01",
  "range": {"lower": {"open": true},
            "upper": {"key": 10, "inclusive": true}},
  "selected": [1, 2, 3, 5, 6, 8, 9, 10],
  "changed": [9]
}
\end{verbatim}
\end{minipage}\par\vspace{2.85pt}
\end{trivlist}
\end{footnotesize}
The selected and changed keys can each be encoded as a sorted list of keys or,
for integer keys, as a list of key intervals.

\textbf{Scope.} The chunk info record is designed for a sink that consumes one
ordered stream. On a key-partitioned transport with several partitions, such as a
Kafka topic, the connector instead can emit individual per-key absence events
(Section~\ref{sec:absence}).

\textbf{Semantics and emission.} At the chunk's close, the connector emits the chunk info record first, followed by the surviving 
chunk rows, so that the sink receives the range bounds and both key lists before processing the chunk rows. 
Change events are emitted like in the original DBLog algorithm. Every event logged up to the high watermark, the window's events included, 
is emitted before the record and every later event after the chunk rows. Every key within the specified range that is neither selected nor 
changed is determined as absent at the source and a sink applying the record deletes every such key it holds. Both the chunk's range bounds 
and the end-of-table test described next are determined from the raw result set of the \texttt{SELECT} query before any rows of changed 
keys are dropped.

\textbf{Covering the entire key space.} Consecutive chunk ranges cover the entire key
space of the table without gaps. The first chunk's range is open at the lower end and each subsequent chunk's range starts after the previous chunk's upper
bound (exclusive), normally its last raw key, and extends up to its own last
raw key (inclusive). A chunk's upper
bound is left open only when the \texttt{SELECT} query returns fewer rows than
the requested chunk limit, which serves as the terminal test indicating the end
of the table. Otherwise, the range closes at the last returned key, where the
next chunk query resumes. Without this rule, a full chunk page at the end of a
table would falsely certify all higher keys as absent, including rows inserted
above that key while the pass was running. While classic DBLog simply scans chunk after chunk until a query returns a short
page, connectors such as Debezium query the table's maximum key once at the
start of a snapshot pass and bound the main chunk loop at that recorded maximum~\cite{debezium_incremental_snapshots}.
When a chunk query bounded this way returns a short page, it closes its range at
that recorded maximum (inclusive) rather than leaving the upper bound open,
because the query was stopped by the bound and not by the end of the table, and
the first tail read starts after that maximum.
Since new rows may be inserted above that initial maximum while the pass is
running, the bounded loop is followed by one or more unbounded \emph{tail
reads} above the recorded maximum that follow the same short-page rule up to a
configured limit (Section~\ref{sec:subtleties}). A tail read that returns a
short page has reached the end of the table, so its range is marked as open at
the upper end to cover all higher keys, completing the pass.

\begin{figure*}[t]
\centering
\includegraphics[trim=0 2.5pt 0 2.5pt,clip]{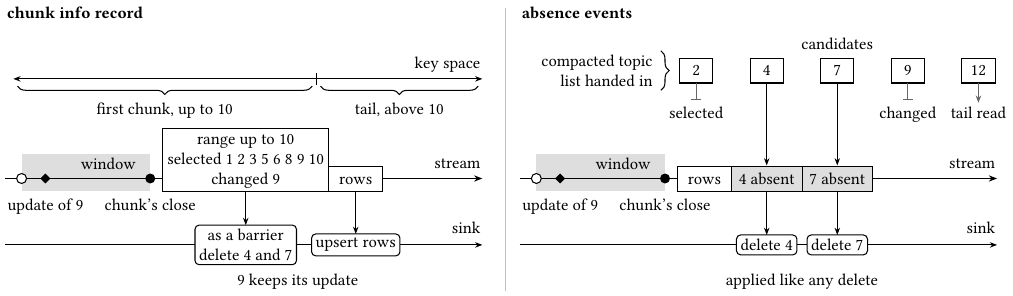}
\caption{The two forms of the absence statement at the chunk's close. On the
left, the chunk info record goes out ahead of the rows and specifies the range up to 10 with its selected and changed keys, and the sink deletes every other key
it holds in that range before applying the rows, while consecutive chunk ranges
cover the key space without gaps along the top strip. On the right, per-key rendering as
absence events removes the selected key 2 and the changed key 9 from the
candidate keys (here a submitted list), leaving keys 4 and 7 to receive absence
events while key 12 waits for the tail read.}
\label{fig:side}
\end{figure*}

\textbf{Empty tables.} An empty source table produces a single chunk info
record with both bounds open and zero rows, instructing the sink to remove
every row it holds for the table except the keys on the record's changed list
(subject to the safety caps of Section~\ref{sec:subtleties}). Because consecutive chunk ranges leave
no gaps across the key space, every orphan key not marked as changed in its
chunk info record is removed once that record is emitted and applied,
without requiring candidate key lists or any sink queries by the source
connector (Figure~\ref{fig:side}).

\textbf{Sink application.} A sink connector that applies chunk info records treats
each one as an ordering barrier in the stream. All
events preceding the chunk info record are applied first and no subsequent events
are processed until the range delete completes. The sink deletes every key it
holds within the chunk's bounds that is neither selected nor changed and then applies
the surviving chunk rows that follow as upserts. The sink can execute the range
delete and the following chunk upserts within a single transaction, or
separately, since the range delete never affects any selected key. A single
transaction requires the sink to recognize the chunk's rows, for example by the
read's identifier attached to them, one row for each selected key that is not
changed. In our running example, orders 4 and 7 are deleted while order 9 retains 
the update delivered by the log window.

A sink that versions each row with its output stream position (the message
offset in Kafka) can execute the range delete without a full stream barrier. In
this versioned mode, the sink atomically deletes a row in the range only if its
key is on neither list and the stored row carries a position before the
record's position (or carries no position), while retaining the record as a range tombstone that rejects older
out-of-order events for keys on neither list, even for keys with no stored row,
so keys can be processed in parallel. Stored positions come from
the current incarnation of the stream, never from a source log coordinate or a
side writer. Re-applying a record or replaying the stream at original positions
is idempotent and a re-executed chunk emits a new record at a later stream
position.

\textbf{Applying the record in practice.} A sink that applies chunk info
records removes every key it holds in the record's range that is on neither of
the record's key lists. A store can do that with one conditional delete that
evaluates the range and both lists, or the sink connector can list the keys
the store holds in the range, keep those on neither list and delete them one
by one. Such a listing shows every row the store holds, including those the
sink connector has just written, and the sink finishes the deletes before
moving past the record, including across restarts. Keys are compared under the
source's key order and equality (Section~\ref{sec:subtleties}). A store with an
ordered index lists a range in one scan. A store without one can take one snapshot of all its keys when the pass's
first record arrives and filter it per record, since a snapshot of the sink's
keys taken at any stream position during the pass suffices for coverage
(Section~\ref{sec:checked}). Orphans a side writer adds after that snapshot
wait for a later pass or a repair request. A store that cannot list its keys,
such as a cache with get, set and delete only, cannot apply chunk info records
and no external process can scan it, so it uses per-key absence events
(Section~\ref{sec:absence}) and repair requests (Section~\ref{sec:request})
instead.

\textbf{Tradeoffs.} The chunk info record costs one metadata record per chunk
regardless of how many orphans exist at the sink. In exchange, applying it
calls for evaluating range deletes at the sink under the source's key order and
equality and requires sink connector logic for the new record type, which 
already released sink connectors do not support today.

\subsection{Absence events, the chunk info record rendered per key}\label{sec:absence}

\textbf{Where per-key absence events fit.} Many CDC pipelines publish to key-partitioned
transports such as Kafka topics and feed off-the-shelf sink connectors that only
process individual per-key events. For these pipelines, the source connector can
render each chunk's absence information into individual \emph{absence events},
one per absent key among the candidate keys supplied to it, which sinks apply
as deletes.
Because per-key rendering only emits absence events for candidate keys known to
the connector, an orphan outside that candidate set stays in place, wherever it
lies, unless the external drift process of Section~\ref{sec:drift} finds it at
the sink. Outside the key bounds that the pass publishes at its end, the bounds
hand-off described below also finds such orphans, since the sink can be asked
for every key it holds there. Chunk info records cover the keys below and above those bounds by themselves, since the first range is open below and the last range open
above. Each
output stream that carries absence statements is configured to emit either chunk info records (for sinks that apply them) or per-key absence events (for standard per-key
sinks), while the per-key absence events of a \texttt{verify-keys-absent}
request (Section~\ref{sec:request}) reach both kinds of output, since a sink
that applies chunk info records also applies absence events as deletes.

\textbf{Record format.} DBLog emits surviving chunk rows as \emph{read} events,
stating that a bracketed read found a key present at its close, which sinks
apply as upserts. In the same way, per-key rendering emits each confirmed
absent key as an \emph{absence event} with the operation type \texttt{absent},
stating that the read found the key absent at its close, which sinks apply as
a delete. Distinguishing an \texttt{absent} event from a log \texttt{delete}
event lets consumers tell an absence observation apart from a live deletion at
the source. A consumer that reacts to business deletes treats an
\texttt{absent} event as it treats a \texttt{read} event, as a statement about
the source's state rather than a transaction. Each absence event carries the
primary key, the table name, the operation type and the identifier of the read
that produced it. On Kafka, it is emitted by default as a standard tombstone
record with the metadata attached in message headers, matching the format that
off-the-shelf sink connectors already consume for deletes~\cite{kafka_docs}.

\textbf{Candidate keys.} In the running example the absent keys of the first
chunk are easy to name. The range covers keys up to 10 and the query returned
eight of them, so with keys starting at 1 the absent keys are 4 and 7. For a
small range of integer keys the source connector could list the absent keys
like this and emit one delete for each. This simple view does not hold in general. A range
of UUIDs or of multi-column keys can hold far more possible keys than rows. 
Even an integer range with gaps holds many absent keys that never
existed as rows and listing them all would flood the stream with deletes for
keys no sink ever held. This is why per-key rendering works from a list of
\emph{candidate keys}. These are keys a sink is suspected to contain and one absence event
is emitted for each candidate in the chunk's range that is neither selected nor changed. A chunk info
record needs no such list, because the sink applies the record's rule to the
keys it holds.

The connector obtains this candidate list when starting a pass from a supplier,
such as a plugin registered with the connector or a signal request. For example,
when the connector writes to compacted Kafka topics, the supplier plugin can
build the list from the output topics, which retain every key previously emitted
for the table except those whose tombstones have already been compacted away. At
the start of a snapshot pass, the plugin scans those topics once using an
internal consumer (similar to how Debezium reads its schema-history and signal
topics), without needing any access to the sink. This topic scan catches lost
deletes (whose last known values remain in the topic) as well as restored rows
and skipped records whose tombstones have not yet expired. However, a topic scan
cannot discover keys that the connector never emitted (such as rows inserted by
a side writer) or keys whose tombstones were already removed by compaction and it is only as complete as the topics themselves. The drift process
(Section~\ref{sec:drift}) covers those remaining cases by finding orphans at the
sink and submitting their keys via the repair request
(Section~\ref{sec:request}). Including a live key on the candidate list is
harmless, as long as keys are compared under the source's key order and
equality. Candidates only nominate keys to check. The bracketed read and the
watermark window determine whether a key is actually absent.

\textbf{Emission.} At the chunk's close, the connector takes all
candidate keys that fall within the chunk's key range, removes every
selected key and every changed key and emits one absence event for each remaining candidate
alongside the surviving chunk rows. In our running example, suppose an operator supplied the suspect keys 2, 4, 7, 9 and 12 as the
candidate list when starting the pass. Key 12 is an extra row a side writer
wrote at the sink, so the topics could not have supplied it. Key 2 was selected by the query, key 9 was changed in the
window and key 12 lies above the first chunk's range, leaving keys 4 and 7 to
receive absence events (Figure~\ref{fig:side}). The relative order between the
chunk's surviving rows and its absence events does not matter because their key
sets are disjoint.

\textbf{Head, tail and bounds hand-off.} Candidate keys below the lowest key in
the source table are covered by the first chunk (whose range is open below) and
candidates above the initial maximum key, such as key 12 in the running
example, are covered by the tail reads.

To handle orphan keys outside the source table's minimum and maximum keys
without scanning the whole sink, at the end of a full pass the connector can publish the pass's observed key
bounds to a dedicated channel such as a Kafka topic. Those bounds consist of
the lowest key returned during the pass and the highest key returned at or
below the initial recorded maximum (or that recorded maximum itself if no keys
at or below it were returned), or empty bounds for an empty table. The upper
bound stops at the recorded maximum because rows inserted during the pass
usually arrive above it and a bound at the highest key the tail reads
returned would hide an orphan between the recorded maximum and those new rows
from the handler's scan. After the bounds are published, a handler beside the sink (either part of the
drift process or a standalone consumer) waits until the sink has caught up with
the stream and then queries the sink for
any keys lying outside those published bounds and submits them to the connector
via \texttt{verify-keys-absent}. A scan before that point is still safe
because the connector's confirming read decides at the source, though it can miss an
orphan whose stale upsert is still in flight and leave it for a later scan. On a table where new rows are continuously
inserted above the initial maximum key during the pass, that query at the sink
will also pick up those newly inserted keys, but the connector's confirming read
will find them present at the source and leave them untouched. With this
bounds hand-off, the
source connector still never accesses the sink directly and safety does not
depend on the handler running, since omitting the hand-off merely leaves any
out-of-bounds orphans in place until a later request names them.

\begin{figure*}[t]
\centering
\includegraphics[trim=0 2.5pt 0 2.5pt,clip]{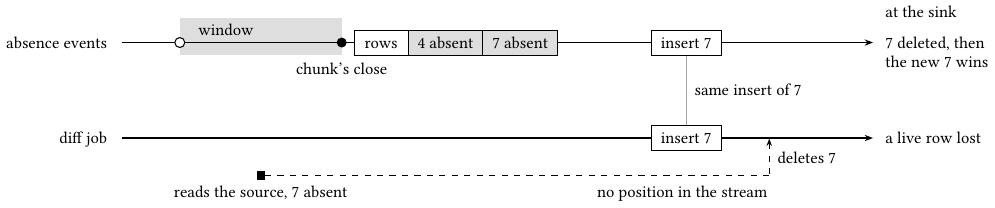}
\caption{Above, the absence event for order 7 sits at the chunk's close, so a
later insert of 7 follows it in the stream and wins at the sink. Below,
the delete of a diff job is decided outside the stream. It has no position and
can land after the pipeline applied the insert, which removes a live row.}
\label{fig:position}
\end{figure*}

\textbf{Stream position.} Because the absence event for order 7 is emitted at
the chunk's close, it sits in the stream after every prior event for key 7 that
took effect before the read and ahead of any subsequent re-insertion of key 7,
which therefore follows the delete in the stream and wins at the sink
(Figure~\ref{fig:position}).

\textbf{Sink configuration.} A sink that consumes per-key absence events
processes them just like ordinary delete events and applies chunk rows as
upserts, which in the Debezium JDBC sink means
\texttt{insert.mode=upsert}~\cite{debezium_jdbc_sink}. For consumers that
already delete rows upon receiving a tombstone, no code changes are required. In popular
Kafka Connect sink connectors, tombstone deletion is a configuration option that
is disabled by default, such as \texttt{delete.enabled} in the Debezium and
Confluent JDBC sinks~\cite{debezium_jdbc_sink,confluent_sinks} and
\texttt{behavior.on.null.values} in the Elasticsearch sink, which defaults to
failing on null payloads~\cite{confluent_sinks}. In addition, any single-message
transforms in the pipeline must be configured not to drop
tombstones~\cite{debezium_event_flattening}. For consumers that parse
operations from a structured CDC envelope rather than raw tombstones, the
connector can emit absence events in that envelope format on a per-output
basis, either with the \texttt{absent} operation type or, for consumers that do not
know the new type, mapped to the \texttt{delete} operation type with the
absence metadata preserved in headers.
In all cases, the message key must be serialized exactly like the key of the
table's ordinary change events, so that the absence event lands in the same
partition and compacts the same key, and an envelope-format absence event on a
compacted topic is followed by a tombstone, as a delete event is, so that the
key leaves the topic. This presumes that the topic's partition count stays fixed, as every
consumer that relies on per-key order already requires.

\textbf{Cost and restarts.} Rendering per-key absence events emits one absence event for every candidate
key in a chunk's range that is neither selected nor
changed, which includes
keys legitimately deleted within the topic's tombstone retention window and can
produce a burst of tombstones following a large batch delete at the source. If
the connector restarts during a snapshot, it simply reloads the candidates (which is safe because emitting a duplicate absence event for an already absent
key is a no-op) and retains any candidate keys submitted via a signal
request until their bracketed read completes.

\subsection{The request verify-keys-absent}\label{sec:request}

While chunk reads check key ranges, a bracketed key-list read answers an
explicit repair request from an external caller. For keys that are present at
the source, DBLog already supports requesting a snapshot for specific
keys~\cite{andreakis2020dblog}, which Debezium provides as an
\texttt{execute-snapshot} signal with an \texttt{additional-conditions}
filter~\cite{debezium_signalling,debezium_postgresql_connector}. Such a
snapshot emits read events for the listed keys it finds present and unchanged
in the window, but emits nothing for the remaining keys and therefore cannot
state absence. Our proposed \texttt{verify-keys-absent} signal request is the
counterpart for absent keys, taking a list of primary keys that the caller
suspects are orphans.
In Debezium's signaling format~\cite{debezium_signalling}, a request to verify
and remove suspect keys 4, 7 and 12 from \texttt{inventory.orders} takes the following form.\par
\begin{footnotesize}
\begin{trivlist}\item\relax
\noindent\begin{minipage}[t]{\columnwidth}
\begin{verbatim}
{
  "id": "repair-orders-01",
  "type": "verify-keys-absent",
  "data": {
    "data-collections": ["inventory.orders"],
    "keys": [4, 7, 12]
  }
}
\end{verbatim}
\end{minipage}\par\vspace{2.85pt}
\end{trivlist}
\end{footnotesize}
When the connector receives this request, it queries those specific keys at the
source between a fresh low and high watermark of its own. At the high watermark,
it emits an absence event for every requested key that was neither returned by
the query nor changed inside the watermark window and it emits no absence event
for keys that are present at the source or were changed in the window (since the
log stream carries the latest state for changed keys). The external requester
never deletes from the sink itself. Any external entity that inspects the sink,
whether a diff job, a bounds handler, or a human operator, can submit the
request for keys anywhere in the key space (the request-driven way in 
Figure~\ref{fig:styles}).

To bound database load, the connector processes requested keys in bounded
batches, executing one bracketed read per batch serialized with ongoing chunk
reads so their watermark windows never overlap. Because the connector emits
absence events only for keys that it confirms absent at the source with an
explicit key-list read, a wrong or malicious request cannot delete a live row
and the resulting absence events propagate to every sink subscribed to the
output. Passing live keys in a request is harmless as long as keys are specified using the
source's key equality.

This signal request is a proposed extension. Debezium's signaling mechanism
currently supports starting, stopping, pausing and resuming snapshots as well
as registering custom actions through its \texttt{SignalActionProvider} SPI, but none of its built-in signals emit events for absent keys~\cite{debezium_signalling}. A
custom signal action can accept \texttt{verify-keys-absent} and reuse
the incremental snapshot engine's watermark windowing to execute the key-list
read.

\textbf{Bridging chunk info records to per-key sinks via a sidecar handler.} Combining
chunk info records (Section~\ref{sec:chunkinfo}) with
\texttt{verify-keys-absent} brings full range coverage
to off-the-shelf sinks that only accept per-key events. In this setup, the source connector publishes chunk info records to a dedicated Kafka topic consumed by
a lightweight handler process running alongside the sink. After it receives a record, this handler waits until the sink has applied the stream up to the point at
which the record was published and those writes are visible to its scan, so
that the scan sees every orphan the record can settle and then queries the sink for any keys present in the
chunk's range that are neither selected nor changed (which catches side-writer keys as well, without needing
pre-existing candidates) and submits those keys back to the source connector
via \texttt{verify-keys-absent}. As with the bounds hand-off of
Section~\ref{sec:absence}, an earlier scan is safe but can miss an orphan whose
stale upsert is still in flight, which a later scan can catch. The handler never writes to the
sink or injects its own delete records into the change stream, because
such records would lack a verified position in the source log. A connector can support both approaches on a
per-output basis, either rendering per-key absence events itself or publishing
chunk info records for an external handler.

\subsection{The drift process and its contract}\label{sec:drift}

A snapshot pass that emits chunk info records covers the entire key space,
including the open regions below the table's lowest key and above its highest
key. Per-key absence events instead check only the candidate key list supplied
to the connector, which may be incomplete, for example missing rows inserted by
a side writer or keys whose tombstones were already compacted away. In
addition, operational considerations can cause a pass under either
representation to stop at the tail-read limit under high write pressure or hold
back a statement at a safety cap (Section~\ref{sec:subtleties}). The external
drift process is designed for these cases. It inspects the sinks and submits
repair requests for the orphans a pass leaves behind.

Together, these mechanisms support three operating modes for orphan removal.
The first mode uses the in-band way of Figure~\ref{fig:styles} and the second
mode the request-driven way, while the out-of-band way is the baseline of
Section~\ref{sec:diff}. In the \emph{fully integrated mode}, every chunk-driven
snapshot pass emits chunk info records or their per-key absence events. When
sinks apply chunk info records, a full snapshot pass removes every orphan across
the completed chunk ranges, repairing a drifted sink without anyone needing to
know which keys were corrupted, which makes it well suited for scheduled
periodic repairs. When sinks consume per-key absence events, a pass repairs the
orphan keys present in its candidate list. In the \emph{drift-driven mode}, an
external drift process monitors the sinks and submits suspected orphans to the
connector via \texttt{verify-keys-absent}. Because the source connector never
reads the sinks, only an external process that inspects a sink can actively
detect that drift has occurred, discover orphans missing from the candidate
list under per-key rendering and catch localized drift much sooner than
periodic full-table passes. The \emph{combined mode} runs both together and is
what we recommend in practice.

We specify the drift process solely by the safety contract it must follow. It
may inspect sinks using any detection method, whether comparing primary keys,
querying a sink for keys outside the connector's published pass bounds, or
running the row counts and range checksums from Section~\ref{sec:kinds}.
However, it must never write directly to a sink and must never publish delete
records into the event stream on its own. Its only output is a
\texttt{verify-keys-absent} request to the source connector. 
Because the connector verifies every
requested key with a bracketed read, the drift process's findings can be
approximate or stale without risking data loss. It can run on a periodic
schedule, on operator demand, or immediately upon receiving a bounds hand-off
from the connector. Any orphans that a partial drift check misses simply remain
until covered by a subsequent request or full pass
(Section~\ref{sec:checked}). The drift process requires only read access to the stores it inspects and
write access to the connector's signal channel. In fact, the external
diff job described in Section~\ref{sec:diff} becomes a compliant drift process
as soon as it replaces its direct sink deletes with repair requests to the
connector.

When a pipeline combines chunk-level absence statements, the end-of-pass bounds
hand-off and a drift process submitting repair requests, every sink that
applies these statements converges to the source state across all covered keys,
under the assumptions of Section~\ref{sec:checked}. A full source-to-sink data
comparison then serves a complementary role as an audit check verifying that a
sink matches the source as of a specific log position already applied by the
sink. Designing a live cross-system comparison against a continuously mutating
source is outside the scope of this paper.

\subsection{Operational considerations and engine subtleties}\label{sec:subtleties}

\textbf{Adaptation to Flink CDC's range merge.} Flink CDC's incremental snapshot
design fits this model with a specific adaptation. Flink CDC reads each chunk
over a pre-declared key range, merges the watermark window's log changes
directly into the buffered chunk image and does not separately forward log
changes for the chunk's keys that occurred before the chunk's close. In our
terms, its effective window at release has zero width, no key counts as changed,
a statement's key range is the declared split range with its own bounds (never
derived from the \texttt{SELECT} rows or left open at the top because of a
partial page), and the keys present in the merged image play the role of the
selected keys wherever our mechanisms refer to them. Because Flink CDC does not
forward pre-close log changes that its query missed, it requires a chunk read
that sees every change logged before it. In addition, because Flink CDC holds back the log changes for a chunk's keys
until the chunk closes, folding the window's changes into the buffered chunk
image and dropping earlier ones that the chunk read already reflects, a key
deleted by a held-back log event is simply absent from the merged image and
no delete event reaches the sink. At the chunk's close, the connector must therefore emit an ordinary
delete for every key in the chunk's range that was deleted by a held-back log
event (whether before or inside the window) and remains absent from the final
merged image, even if the chunk's statement of absence is withheld or refused.
To catch deletes that occurred before the chunk's window, the connector must
process the table's log continuously from the start of the pass and keep the
keys of each split's held-back changes until that split closes.

\textbf{Chunk query visibility.} Every statement of absence assumes that the
chunk query sees every transaction that precedes the low watermark in the log.
DBLog's window rule handles everything after that point, because a transaction
logged between the two watermarks marks its keys as changed and the log rather
than the chunk row decides their state. If that assumption fails, a row
inserted by a transaction that reached the log before the low watermark but was
not yet visible to the query is neither selected nor changed, so the statement
certifies it absent and a sink applying it deletes a live row. PostgreSQL is an
engine where a committed transaction can appear in the log before it becomes visible to a new query, briefly under
its default commit behavior and, under synchronous replication, until a standby
confirms the commit~\cite{postgres_xact_source}. On such an engine the connector treats the keys of every transaction logged
before the chunk's close that its query did not see as changed, which excludes
them from the statement and leaves their state to the log. Every statement then
stays safe, but the pass copies no row for such a key, so a sink that lost the
transaction's events before the pass started keeps lacking the row until a
later read copies it. The connector tells those transactions apart by
capturing the query's own snapshot and checking each decoded transaction
against it. Waiting for the transactions logged before the low watermark to
finish before running the query avoids the gap altogether and the convergence
results of Section~\ref{sec:checked} assume reads of that kind. MySQL in its
default configuration does not have this gap, since transactions commit in
binary log order and a transaction that precedes the low watermark in the log
has already committed when the watermark write
returns~\cite{mysql_binlog_expire}. The bracketed key-list read of
Section~\ref{sec:request} applies the same rule with its own window.

\textbf{Key ordering and equality.} Every statement of absence is defined under
the source database's key ordering and key equality. A sink that applies a
statement evaluates range bounds, membership in the selected and changed lists,
and per-key deletes under that same ordering and equality. If a sink compares strings byte-wise while the source
uses a case-insensitive collation, a key can fall into the wrong range and be
deleted while still live at the source. One practical way to guarantee matching
semantics is for the source connector to ship an exact, order-preserving byte encoding
for every key, bound and list entry so that the sink performs plain byte
comparisons. Data types that lack such an encoding (such as floating-point keys with special
values) and tables chunked by a surrogate key not stored at the sink cannot use
chunk info records. A connector is therefore safest when it enables chunk info
records only for key types it can encode exactly and in the source's order.

\textbf{Tail-read limit under high write pressure.} On a table that
continuously receives new inserts above the maximum key recorded at the start
of a pass, the tail reads of Section~\ref{sec:chunkinfo} can reach their
configured limit before seeing a short page, leaving the remaining key space
above the last returned key uncovered by a chunk info record. Under per-key
rendering, the connector still checks any candidate keys in that remaining tail
using bracketed key-list reads at the end of the pass, while the drift process
of Section~\ref{sec:drift} (or a later pass) catches any other orphans there.

\textbf{Safety caps.} To prevent accidental mass deletions, a configurable
safety cap applies to both representations, and an empty source table, which a pass recognizes when its first chunk read
returns no rows, always calls for operator confirmation. For chunk info records, a sink can refuse an
unconfirmed empty-table record or any record that would delete more rows than
the cap. For per-key absence events and \texttt{verify-keys-absent} requests,
the source connector holds back and reports any chunk or request that exceeds
the cap (and always an empty table). Once an operator confirms via a signal,
the connector runs a fresh bracketed read and emits the events or a confirmed
chunk info record.

\textbf{Table-level changes and restarts.} A table-level change is any change
to the rows or keys a chunk query reads that the log does not carry as row
events. Any such change inside the window, such as a \texttt{TRUNCATE},
\texttt{DROP}, \texttt{RENAME}, or a modification to primary-key columns or
their collation, aborts the chunk's statement of absence and forces the chunk to be re-read in a
fresh window, just as an open chunk interrupted by a connector restart is
discarded and re-read. A chunk stays open until all events of its close
are committed. A \texttt{TRUNCATE} or a \texttt{DROP} also invalidates
the chunks completed before it, since the rows they copied are now orphans that
no row event mentions, so the pass restarts from its first chunk unless the
pipeline applies the table-wide change at its position in the stream.

\textbf{Dead-letter queues, archives and multi-source sinks.} On the consumer
side, a failed record that was routed to a dead-letter queue must never be
replayed on its own later, because the sink has already advanced past that
stream position. Only replaying the entire stream in order from that position is
safe, unless the sink tracks the highest stream position per row, restricts
every applied operation by it (Section~\ref{sec:chunkinfo}), and the dead-letter
record keeps its original stream position. Sinks that
intentionally retain rows deleted at the source, such as historical archives, or
tables populated by merging multiple sources should not apply chunk info
records or absence events. They should either consume an output stream configured without
absence statements or filter them out using the origin metadata in the record
headers.

\section{The baseline, a diff job}\label{sec:diff}

A standalone diff-and-delete job requires no changes to the CDC connector. It
reads the primary keys from both the source and the sink and directly deletes
from the sink any key missing at the source (orders 4, 7 and 12 in our
example). Commercial replication tools such as Oracle GoldenGate Veridata and
AWS DMS data resync provide this capability, coordinated with replication, for
example by confirming out-of-sync rows after a delay for replication lag or by
pausing replication while fixes are applied~\cite{goldengate_veridata,aws_dms_resync}.
A standalone job that executes deletes outside the CDC stream without such
coordination introduces two fundamental hazards.

\textbf{No position in the stream.} Because every write or delete performed by
the diff job happens outside the ordered CDC log stream, a direct delete can
remove a live row and an out-of-band upsert can overwrite a newer column value
with an older one.

\textbf{Data in flight and the ordering condition.} The diff job queries the
source at one instant and issues its delete against the sink at a later instant.
Suppose the application re-creates order 7 at the source in the meantime (as can 
happen with natural primary keys or re-inserted records) and the CDC
pipeline delivers that new insert to the sink before the diff job's delete
arrives. Because nothing orders the out-of-band delete relative to the stream's
insert, the diff job deletes the newly created live row
(Figure~\ref{fig:position}). Conversely, if the diff job's delete lands at the
sink first and the stream's insert arrives second, a sink that retains the latest
write converges to the correct state.

More precisely, an out-of-band delete for a key is safe under the following
\emph{ordering condition}. First, the diff job's read must observe the key as
absent at some source log position before which all committed changes were
visible to that read (the counterpart of the visibility assumption of Section~\ref{sec:subtleties}).
Second, at the moment the delete takes effect at the sink, the sink must not yet
have applied any change event for that key originating after that source log
position (and on a sink that resolves conflicts by keeping the highest version,
the delete must carry a version that sorts above the key's earlier events and
below its later ones). If the sink is lagging
behind the diff job's read position when the delete runs, no live row is lost,
although an older change event for that key still in flight in the pipeline can
re-create the orphan if its subsequent delete was lost.

Our in-band mechanisms satisfy this ordering condition by construction, because
the watermark window covers the entire interval between the source read and the
emission of the delete at the chunk's close. An external diff job can satisfy the condition in three ways. It can freeze
writes to the source table from its read until its sink deletes complete. It can
first note how far the stream that feeds the sink has advanced, per partition
on a partitioned topic, then take a fresh source snapshot for its read that sees every transaction
logged before the positions it noted (the visibility assumption of
Section~\ref{sec:subtleties}) and then execute a conditional delete at the
sink that only removes rows whose
stored stream position lies before the position noted for their partition. Or
it can pause sink event application for the suspect key while re-verifying its
absence at the source before deleting. Alternatively, the
diff job can simply hand the suspect keys to the connector via
\texttt{verify-keys-absent} (Section~\ref{sec:request}), letting the connector verify them and assign each absence event a proper
position in the stream.
Note that publishing a tombstone directly into the connector's Kafka topic from
an external job does not solve the problem, because that tombstone is still not
ordered against the source database's transaction log.

\textbf{Operational cost.} Because each run requires full key scans of both the
source and the sink, teams typically run full diff jobs infrequently, leaving
orphan rows in the sink between runs.

\section{Comparing the three approaches}\label{sec:compare}

\begin{table*}[t]
\caption{The three approaches side by side.}
\label{tab:compare}
\footnotesize
\setlength{\tabcolsep}{4pt}
\renewcommand{\arraystretch}{1.1}
\begin{tabularx}{\textwidth}{@{}>{\RaggedRight}p{1.0in}>{\RaggedRight}p{1.62in}YY@{}}
\toprule
  & Diff job & Chunk info record & Absence events \\
\midrule
What changes & Nothing in the pipeline, a new external job & The source connector and a sink
connector component that applies chunk info records & The source connector, plus enabling delete-on-tombstone
at the sink \\
\addlinespace[2pt]
Who deletes & The diff job, directly at the sink & The sink, by key range & The sink, by key, just like an ordinary delete \\
\addlinespace[2pt]
Which orphans are found & Every orphan present at the time of comparison & Every orphan in the applied ranges not marked as changed & Candidate keys supplied to the connector, such as from its own topics, plus keys submitted via repair requests \\
\addlinespace[2pt]
Position in the stream & No & Yes & Yes \\
\addlinespace[2pt]
Cost & Two full scans per run & A snapshot pass, one chunk info record per chunk
and range evaluation at the sink & A snapshot pass, a topic scan and one event per candidate key found absent \\
\addlinespace[2pt]
Runs while data flows & Only under the ordering condition & Yes & Yes \\
\bottomrule
\end{tabularx}
\end{table*}

Table~\ref{tab:compare} summarizes the tradeoffs across the three approaches. No
single option dominates in every deployment. Between the two connector-side
representations, per-key absence events require minimal changes at the sink
(often just flipping a configuration flag) but only cover candidate keys supplied to the connector, from its own topics or
a repair request. Chunk
info records call for a sink that removes keys in a range, but they achieve
full key-space coverage without needing candidate lists. A connector can also emit
both representations on separate outputs and the
\texttt{verify-keys-absent} request allows an external drift process to
clean up any orphans that either representation left behind (Section~\ref{sec:drift}).

Under both connector-side representations, any key that happens to be changed
inside the watermark window of every snapshot pass is deferred to the log stream
rather than settled by a chunk read.

\section{What is formally verified}\label{sec:checked}

Race conditions between live log events, chunk reads and deletes depend on
exact event interleavings that are easy to miss in testing. To rule out
protocol-level bugs, we modeled our extensions on top of the Generalized DBLog
formal model~\cite{andreakis2026generalized,generalized_dblog_artifact} and
mechanically verified their safety and convergence across all possible event
histories and initial sink states in Isabelle/HOL, with the main results proved
again in Lean~4~\cite{orphan_absence_artifact}. The model represents a sink
that starts from an arbitrary state (containing any mix of missing rows, stale
values and orphans) and applies the emitted event stream in order. It
establishes three guarantees.

\textbf{Safety (no false deletes or wrong values).} Emitting surviving chunk
rows together with deletes only for keys found absent by the read and unchanged
inside the watermark window never writes a wrong value and never deletes a
live row. Any orphan that neither a read nor the log stream mentions simply
remains untouched.

\textbf{Convergence (exact synchronization).} A sink is guaranteed to match the
exact source state at the end of a pass if and only if the emitted absence
statements cover every orphan held by the sink that the log stream never
mentions. We prove that this holds when applying chunk info records across the
table's ranges, when per-key absence events are given a complete snapshot of
the sink's keys taken during the pass, and when a pass that left orphans behind is followed by a
\texttt{verify-keys-absent} request (or a later pass) that checks those
remaining keys.

\textbf{Flink CDC range-merge rule.} For Flink CDC's buffered range merge, which
holds back log changes for a chunk's keys until the chunk closes, the proofs
show that omitting deletes for keys removed by held-back events leaves a
deleted row behind in the sink, and that re-emitting those deletes at the
chunk's close (Section~\ref{sec:subtleties}) gives the same safety and
convergence.

\textbf{Further results.} We also verified four operational cases. A repair request served while a pass is still running
keeps safety and convergence, provided its window lies within the pass. The same holds at the end of a resumed pass when one connector restart
interrupts a pass outside a chunk's close, provided the connector resumes from
an offset committed during the pass and re-runs every unfinished chunk. The chunk ranges of a pass never overlap and leave no gaps, covering
every key when the pass completes (or all keys up to the last returned key if
stopped by the tail-read limit), and every unselected key in a range is confirmed
absent at that read. Finally, safety and convergence still hold at the end of the stream when the pipeline
delivers each key's events in order at least once rather than exactly once, as a multi-partition topic keyed by primary key does when compaction does not
affect the read.

\textbf{System assumptions and scope.} These proofs verify the stream-level
algorithm. In a deployment, they rely on the pipeline delivering each key's events in
order, enforcing the chunk query visibility rule of
Section~\ref{sec:subtleties}, logging full row images, running complete SQL
\texttt{SELECT} scans, matching key order and equality between source and sink
under a fixed schema, and on transaction atomicity and liveness. Concurrent side writers during repair, row-versioned conflict resolution
(Section~\ref{sec:chunkinfo}) and out-of-order delivery of one key's events
are outside the formal model. All proofs are complete without custom axioms
(with every theory except the Flink CDC range merge proved in both Isabelle/HOL
and Lean~4) and are included in the Zenodo artifact~\cite{orphan_absence_artifact}.
The artifact also includes bounded TLA\textsuperscript{+} models, checked with
the TLC model checker, that test the connector's step-by-step execution,
including crashes at any point, the commit visibility gap and repair requests.
Removing a rule the models test, such as the visibility rule or keeping a chunk
open until all events of its close are committed, makes the model checker
produce a counterexample.

\section{Prior art}\label{sec:prior}

Deleting target rows that are absent from a source block is a long-standing idea
in database replication, documented for Gravic Shadowbase's SOLV utility by
2007~\cite{availability_digest_2007}. SOLV locks and reads a block of rows at the source and injects a marker into the replication stream. In its verification mode the target consumer then reads the corresponding block from the target and compares the two, and in its resynchronization mode it also deletes any target row missing from the source block.
Unlike SOLV as documented in 2007, our mechanisms require no source locks, exclude keys changed inside the watermark window from every
statement of absence and never read or compare row contents at the sink during
a chunk pass. We do not claim novelty for partitioning a key space into contiguous
block ranges (including the open regions below the first block and above the
last), nor for issuing per-key deletes for keys absent from a fresh source read
or executing range deletes at a sink.

Commercial compare-and-repair tools such as Oracle GoldenGate Veridata and AWS
DMS data resync repair out-of-sync rows identified by a completed comparison,
including target rows that no longer exist at the
source~\cite{goldengate_veridata,aws_dms_resync}. Unlike those tools, a
statement of absence from a chunk read in our design is derived directly from
the source connector's own watermark-bracketed query while live changes continue
to flow, takes its exact position at the chunk's close in the CDC stream and
relies solely on that source read. Our \texttt{verify-keys-absent}
request shares their detect-then-confirm pattern, but executes the confirmation
as a watermark-bracketed read inside the source connector so that each
confirmed absence receives a valid position in the change stream. Outside of log-based
CDC, the refresh-present phase of LDAP content synchronization lists the directory
entries that remain present so that clients can delete unlisted unchanged
entries, operating without chunks or watermark windows~\cite{rfc4533}.

Compared to existing DBLog-based implementations in Debezium's incremental
snapshots~\cite{debezium_incremental_snapshots} and Flink
CDC~\cite{flinkcdc_mysql}, whose chunk reads emit nothing for a key that the query does not return and that
is not changed in the window, our design retains the absent half of each bracketed chunk read
as either a chunk info record or per-key absence events placed at the chunk's close. In particular, our chunk info record is certified by watermarks at a
well-defined position in the live CDC stream and is applied by the sink without
reading target blocks or performing a confirmation round trip (unless a safety
cap is triggered). This distinguishes it both from range deletes that a full pass or a recovery
session drives without concurrent log streaming and from range repairs that
lock the source, read the target block, wait for a confirmation from the
source, or truncate and reload the entire range.

\section{Conclusion}\label{sec:future}

DBLog's original incremental snapshot algorithm uses only half of the
information established by each bracketed chunk read, emitting the rows that
exist at the source while discarding the fact that all other keys in the queried
range are absent. As a result, existing DBLog-based pipelines could bootstrap
empty sinks and repair missing rows or stale column values, but could not remove
orphan rows left behind by lost deletes, sinks restored from older backups,
skipped delete records, expired tombstones, or side writers. Teams facing such drift previously had to either provision
a new sink from scratch and coordinate a cutover, or rely on out-of-band repair tools that compare source and sink outside the
stream and either pause replication, lock the source, or risk removing live
rows with unpositioned deletes.

By retaining the absent half of each chunk read across chunk ranges that cover the entire key
space and complementing it with the \texttt{verify-keys-absent}
repair request for externally detected drift, our extensions turn DBLog into a
complete data synchronization engine. A sink can now be brought into sync and
kept in sync in place over the existing CDC pipeline whether it catches up on
live changes or missing rows, holds wrong
column values, or contains orphan rows that no
longer exist at the source. This enables both on-demand recovery after an
incident and periodic proactive repairs without interrupting live streaming or
switching to a new sink and for production deployments we recommend combining
chunk-driven passes with an external drift process that submits suspect orphans 
through the connector.

\section*{AI assistance disclosure}

The author used Claude Fable 5.1, Claude Opus 5.5, ChatGPT Astra 6 and
Gemini 4 Argon to assist with the theory, the Isabelle/HOL and Lean~4
formalizations, the figures and drafting and reviewing the paper's prose. The
author edited all of the prose and takes responsibility for the paper's content
and conclusions.

\bibliographystyle{unsrtnat}
\bibliography{refs}

\begin{thebibliography}{27}
\providecommand{\natexlab}[1]{#1}
\providecommand{\url}[1]{\texttt{#1}}
\expandafter\ifx\csname urlstyle\endcsname\relax
  \providecommand{\doi}[1]{doi: #1}\else
  \providecommand{\doi}{doi: \begingroup \urlstyle{rm}\Url}\fi

\bibitem[Kleppmann et~al.(2019)Kleppmann, Beresford, and
  Svingen]{kleppmann2019olep}
Martin Kleppmann, Alastair~R. Beresford, and Boerge Svingen.
\newblock Online event processing: Achieving consistency where distributed
  transactions have failed.
\newblock \emph{Communications of the {ACM}}, 62\penalty0 (5):\penalty0 43--49,
  2019.
\newblock \doi{10.1145/3312527}.

\bibitem[Laigner et~al.(2021)Laigner, Zhou, Salles, Liu, and
  Kalinowski]{laigner2021microservices}
Rodrigo Laigner, Yongluan Zhou, Marcos Antonio~Vaz Salles, Yijian Liu, and
  Marcos Kalinowski.
\newblock Data management in microservices: State of the practice, challenges,
  and research directions.
\newblock \emph{Proceedings of the {VLDB} Endowment}, 14\penalty0
  (13):\penalty0 3348--3361, 2021.
\newblock \doi{10.14778/3484224.3484232}.

\bibitem[Andreakis and Papapanagiotou(2020)]{andreakis2020dblog}
Andreas Andreakis and Ioannis Papapanagiotou.
\newblock {DBLog}: A watermark based change-data-capture framework, 2020.
\newblock URL \url{https://arxiv.org/abs/2010.12597}.
\newblock arXiv:2010.12597v1, 23 October 2020.

\bibitem[Cranford(2021)]{debezium_16_final}
Chris Cranford.
\newblock {Debezium} 1.6.0.{Final} released.
\newblock Debezium Blog, 30 June 2021,
  \url{https://debezium.io/blog/2021/06/30/debezium-1-6-final-released/}, 2021.
\newblock Accessed 2 November 2026.

\bibitem[Pechanec(2021)]{debezium_incremental_snapshots}
Jiri Pechanec.
\newblock Incremental snapshots in {Debezium}.
\newblock Debezium Blog, 7 October 2021,
  \url{https://debezium.io/blog/2021/10/07/incremental-snapshots/}, 2021.
\newblock Accessed 2 November 2026.

\bibitem[{Apache Flink CDC Community}(2026)]{flinkcdc_mysql}
{Apache Flink CDC Community}.
\newblock {MySQL} {CDC} connector.
\newblock Apache Flink CDC documentation (stable),
  \url{https://nightlies.apache.org/flink/flink-cdc-docs-stable/docs/connectors/flink-sources/mysql-cdc/},
  2026.
\newblock Flink CDC 3.7.0 documentation, accessed 2 November 2026.

\bibitem[Andreakis(2026{\natexlab{a}})]{andreakis2026generalized}
Andreas Andreakis.
\newblock Generalized {DBLog}: A verified contract for interleaving copied rows
  with a change log, 2026{\natexlab{a}}.
\newblock URL \url{https://arxiv.org/abs/2609.08160}.
\newblock arXiv:2609.08160v4, 5 October 2026.

\bibitem[Andreakis and Papapanagiotou(2019)]{dblog_netflix_blog}
Andreas Andreakis and Ioannis Papapanagiotou.
\newblock {DBLog}: A generic change-data-capture framework.
\newblock Netflix Technology Blog, 17 December 2019,
  \url{https://netflixtechblog.com/dblog-a-generic-change-data-capture-framework-69351fb9099b},
  2019.
\newblock Accessed 2 November 2026.

\bibitem[Andreakis(2026{\natexlab{b}})]{generalized_dblog_artifact}
Andreas Andreakis.
\newblock Formal development for ``{Generalized DBLog: A Verified Contract for
  Interleaving Copied Rows with a Change Log}''.
\newblock Zenodo, \url{https://doi.org/10.5281/zenodo.22643866},
  2026{\natexlab{b}}.
\newblock Isabelle/HOL, Lean~4 and TLA\textsuperscript{+} sources.

\bibitem[{Debezium Community}(2026{\natexlab{a}})]{debezium_signalling}
{Debezium Community}.
\newblock Sending signals to a {Debezium} connector.
\newblock Debezium reference (stable),
  \url{https://debezium.io/documentation/reference/stable/configuration/signalling.html},
  2026{\natexlab{a}}.
\newblock Debezium 3.7 reference documentation, accessed 2 November 2026.

\bibitem[{Debezium Community}(2026{\natexlab{b}})]{debezium_mysql_connector}
{Debezium Community}.
\newblock {Debezium} connector for {MySQL}.
\newblock Debezium reference (stable),
  \url{https://debezium.io/documentation/reference/stable/connectors/mysql.html},
  2026{\natexlab{b}}.
\newblock Debezium 3.7 reference documentation, accessed 2 November 2026.

\bibitem[Galieva(2022)]{debezium_readonly_blog}
Kate Galieva.
\newblock Read-only incremental snapshots for {MySQL}.
\newblock Debezium Blog, 7 April 2022,
  \url{https://debezium.io/blog/2022/04/07/read-only-incremental-snapshots/},
  2022.
\newblock Accessed 2 November 2026.

\bibitem[{Debezium
  Community}(2026{\natexlab{c}})]{debezium_postgresql_connector}
{Debezium Community}.
\newblock {Debezium} connector for {PostgreSQL}.
\newblock Debezium reference (stable),
  \url{https://debezium.io/documentation/reference/stable/connectors/postgresql.html},
  2026{\natexlab{c}}.
\newblock Debezium 3.7 reference documentation, accessed 2 November 2026.

\bibitem[{Apache Kafka Community}(2026{\natexlab{a}})]{kafka_docs}
{Apache Kafka Community}.
\newblock {Kafka} documentation: Design, log compaction, and topic-level
  configs.
\newblock Apache Kafka 4.4 documentation,
  \url{https://kafka.apache.org/44/design/design/\#log-compaction} and
  \url{https://kafka.apache.org/44/configuration/topic-configs/},
  2026{\natexlab{a}}.
\newblock Accessed 2 November 2026.

\bibitem[{Oracle Corporation}(2026{\natexlab{a}})]{mysql_binlog_expire}
{Oracle Corporation}.
\newblock {MySQL} 8.4 reference manual: Binary logging options and variables.
\newblock
  \url{https://dev.mysql.com/doc/refman/8.4/en/replication-options-binary-log.html},
  2026{\natexlab{a}}.
\newblock Accessed 2 November 2026.

\bibitem[{PostgreSQL Global Development
  Group}(2026)]{postgres_replication_slots}
{PostgreSQL Global Development Group}.
\newblock {PostgreSQL} 18 documentation: Replication (19.6) and logical
  decoding concepts (47.2).
\newblock
  \url{https://www.postgresql.org/docs/18/runtime-config-replication.html} and
  \url{https://www.postgresql.org/docs/18/logicaldecoding-explanation.html\#LOGICALDECODING-REPLICATION-SLOTS},
  2026.
\newblock Accessed 5 November 2026.

\bibitem[{Apache Kafka Community}(2026{\natexlab{b}})]{kafka_connect_configs}
{Apache Kafka Community}.
\newblock {Kafka Connect} configs.
\newblock Apache Kafka 4.4 documentation,
  \url{https://kafka.apache.org/44/configuration/kafka-connect-configs/},
  2026{\natexlab{b}}.
\newblock Accessed 2 November 2026.

\bibitem[{Percona}(2026)]{percona_pt_table_checksum}
{Percona}.
\newblock {Percona Toolkit} documentation: pt-table-checksum.
\newblock
  \url{https://docs.percona.com/percona-toolkit/pt-table-checksum.html}, 2026.
\newblock Accessed 5 November 2026.

\bibitem[{Debezium Community}(2026{\natexlab{d}})]{debezium_jdbc_sink}
{Debezium Community}.
\newblock {Debezium} connector for {JDBC}.
\newblock Debezium reference (stable),
  \url{https://debezium.io/documentation/reference/stable/connectors/jdbc.html},
  2026{\natexlab{d}}.
\newblock Debezium 3.7 reference documentation, accessed 2 November 2026.

\bibitem[{Confluent, Inc.}(2026)]{confluent_sinks}
{Confluent, Inc.}
\newblock {JDBC} sink and {Elasticsearch} sink connector configuration
  properties.
\newblock
  \url{https://docs.confluent.io/kafka-connectors/jdbc/current/sink-connector/sink_config_options.html}
  and
  \url{https://docs.confluent.io/kafka-connectors/elasticsearch/current/configuration_options.html},
  2026.
\newblock Current documentation, accessed 2 November 2026.

\bibitem[{Debezium Community}(2026{\natexlab{e}})]{debezium_event_flattening}
{Debezium Community}.
\newblock New record state extraction ({Debezium} transformation).
\newblock Debezium reference (stable),
  \url{https://debezium.io/documentation/reference/stable/transformations/event-flattening.html},
  2026{\natexlab{e}}.
\newblock Debezium 3.7 reference documentation, accessed 2 November 2026.

\bibitem[{PostgreSQL Global Development Group}(2025)]{postgres_xact_source}
{PostgreSQL Global Development Group}.
\newblock {PostgreSQL} 18.0 source code, file
  src/backend/access/transam/xact.c.
\newblock Functions CommitTransaction and RecordTransactionCommit, release tag
  REL\_18\_0,
  \url{https://github.com/postgres/postgres/blob/REL_18_0/src/backend/access/transam/xact.c},
  2025.
\newblock Accessed 7 November 2026.

\bibitem[{Oracle Corporation}(2026{\natexlab{b}})]{goldengate_veridata}
{Oracle Corporation}.
\newblock Using {Oracle GoldenGate Veridata} 12c (12.2.1.4): Working with jobs.
\newblock
  \url{https://docs.oracle.com/en/middleware/goldengate/veridata/12.2.1.4/gvdug/working-jobs.html},
  2026{\natexlab{b}}.
\newblock Accessed 2 November 2026.

\bibitem[{Amazon Web Services}(2026)]{aws_dms_resync}
{Amazon Web Services}.
\newblock {AWS Database Migration Service} user guide: Data resync and data
  validation.
\newblock
  \url{https://docs.aws.amazon.com/dms/latest/userguide/CHAP_Validating.DataResync.html}
  and
  \url{https://docs.aws.amazon.com/dms/latest/userguide/CHAP_Validating.html},
  2026.
\newblock Accessed 2 November 2026.

\bibitem[Andreakis(2026{\natexlab{c}})]{orphan_absence_artifact}
Andreas Andreakis.
\newblock Formal development for ``{Extending DBLog-Based Incremental Snapshots
  to Handle Orphan Rows in CDC Pipelines}''.
\newblock Zenodo, \url{https://doi.org/10.5281/zenodo.24035424},
  2026{\natexlab{c}}.
\newblock Isabelle/HOL (\mbox{Isabelle2025-2}), Lean~4 and
  TLA\textsuperscript{+} sources, builds on the Generalized DBLog contract
  core.

\bibitem[{Sombers Associates and W. H.
  Highleyman}(2007)]{availability_digest_2007}
{Sombers Associates and W. H. Highleyman}.
\newblock {Shadowbase}: The active/active solution.
\newblock The Availability Digest, March 2007,
  \url{https://availabilitydigest.com/public_articles/0203/shadowbase.pdf},
  2007.
\newblock Accessed 2 November 2026.

\bibitem[Zeilenga and Choi(2006)]{rfc4533}
Kurt Zeilenga and Jong~Hyuk Choi.
\newblock The lightweight directory access protocol ({LDAP}) content
  synchronization operation.
\newblock RFC 4533, June 2006, \url{https://www.rfc-editor.org/rfc/rfc4533},
  2006.
\newblock Accessed 2 November 2026.

\end{thebibliography}

\end{document}